\documentclass[twocolumn,aps,prb,floats,showpacs]{revtex4}
\newcommand{\beq}{\begin{eqnarray}}
\newcommand{\eeq}{\end{eqnarray}}

\usepackage{epsfig}
\usepackage{dcolumn}
\usepackage{bm}

\begin{document}
\bibliographystyle{prsty}
\title{Towards Classification of  Phase Transitions in Reaction--Diffusion Models}
\draft
\author{Vlad Elgart and Alex Kamenev}

\affiliation{Department of Physics, University of Minnesota,
Minneapolis, MN 55455, USA}

\date{\today}


\begin{abstract}
    {\rm
    Equilibrium  phase transitions are associated with
    rearrangements of minima of a (Lagrangian) potential. Treatment  of
    nonequilibrium systems requires doubling of  degrees of
    freedom, which  may be often interpreted as a transition from the
    ``coordinate'' - to the ``phase''--space representation. As a result, one has to deal
    with the Hamiltonian formulation of the
    field theory instead of the Lagrangian one.
    We suggest a classification scheme of phase transitions in
    reaction--diffusion models based on the topology of the phase portraits of
    corresponding Hamiltonians. In models with an absorbing state
    such a topology is fully determined by intersecting curves of zero
    ``energy." We identify four families of topologically distinct classes of
    phase portraits stable upon renormalization group transformations.
      }
\end{abstract}

\pacs{64.60.Ak, 05.40.-a, 64.60.Cn, 82.20.-w}
\maketitle



\bigskip


\section{Introduction}

The last decade witnessed the rapid growth of interest in
reaction--diffusion models
\cite{Hinrichsen00,MarroDickman99,Odor04,Lubeck04,CardyTauber,JanssenTauber05,THL05,HH04,Cardy97,BramsonLebowitz91,Mattis98,bAH00,Schutz01,BrayBlythe03,NPRG}.
Such models are employed for a description of phenomena ranging
from kinetics of chemical reactions to the evolution of biological
populations. The subject of particular interest is the description of
dynamical phase transitions in reaction--diffusion kinetics. An
important example  is absorbing phase transitions. Upon such a
transition the system goes from an active (``living'') phase to an
absorbing (``dead'') state with no escape from it.

Grassberger and Janssen \cite{G82,J81} 
realized that many of the absorbing--state transitions  belong to
the same universality class as the directed percolation (DP)
model. Since then, the DP universality class was extensively
studied both analytically and numerically (see
Refs.~[\onlinecite{Hinrichsen00,Odor04,CardyGrassberger85,Harr74,Kinz83,Ligg85,JG96,GM97,Lub05}]).
The DP universality class is extremely robust.  In fact,
exceptions to the DP transitions are rare. However, if the
microscopic dynamics possesses additional symmetries, the
universality class of the transition may be different. For
example, parity conservation (PC) is known to drive the
transition to a new distinct universality class
\cite{Hinrichsen00,Odor04,CardyTauber,Zhon95,CCDDM05}. Recently
other possible universality classes have been studied
\cite{Hinrichsen00,Dorn01,Droz03,Ross00,Wijland02,Janssen87,OK87,JMS04,PHK02,Lub02,Racz02,Lub03,NP04,Lub06,Grassberger06}.
Among them is a pair contact process with diffusion (PCPD), the
critical behavior of which has not yet been described analytically
\cite{HH04,HT97,Carl01,Hinr01a,Park01,Henk01a,Henk01b,Odor02a,Dick02b,Kock02,Odor02c,Paes03b,BSH06,Bark03,PP05b,H05,DCM05}.

Despite extensive accumulated knowledge, it seems that a
guiding principle, allowing one to  distinguish between various types
of the transitions, is still missing. The purpose of this paper is
to suggest a simple scheme, providing at least, an
educated guess regarding the universality class of the
reaction--diffusion model at hand. The scheme is based on the
topology of the phase portraits of the system's Hamiltonian.
Before elaborating on it, let us briefly recall the corresponding
strategy for equilibrium systems.

An equilibrium system may be characterized by an action (energy)
\begin{equation}\label{equilibrium}
  S=\int d^d x \left[D(\nabla q)^2 + V(q)\right]\, ,
\end{equation}
written in terms  of the order parameter $q(x)$ (for simplicity we
restrict ourselves to the one--component order parameter). The
potential function $V(q)$ encompasses  information about possible
phase transitions. Specifically, one monitors the behavior of the {\em
minima} of $V(q)$ as a function of the control parameter to infer
the existence and type of transition.  For example, a wide
class of models  may be described by a potential of the form:
\begin{equation}\label{potential}
  V(q)=h\,q+m\,q^2+u\,q^4\, .
\end{equation}
For $m<0$ the system exhibits a first--order transition when $h$
changes its sign (for $d>1$) and the two minima interchange.  In the
symmetric case, $h=0$, the system may undergo a second--order
transition when the parameter $m$ is swept through zero, so a single
minimum is split in two.  Below the critical dimension $d_c=4$ this
second--order transition is characterized by non--mean--field
critical exponents. To find the exponents one typically employs the
renormalization group (RG) technique. The RG technique treats the coefficients
$m$ and $u$ of the potential (\ref{potential}) as functions of the
running spatial scale. For $\epsilon=d_c-d>0$ the potential scales
towards the non--trivial fixed point potential, $V^*(q)\neq 0$, with
$m^*\sim\epsilon^2$ and $u^*\sim \epsilon$.  Notice that the action
(\ref{equilibrium}) essentially specifies the Lagrangian field
theory.

One may ask whether a similar strategy  exists for non--equilibrium
phase transitions in reaction--diffusion systems. To answer this
question one has to recall that a description of nonequilibrium
systems requires doubling of the degrees of freedom. There are
various manifestations of this statement depending on the specific
context. In quantum kinetics it is known as the Keldysh technique
\cite{Keldysh}. It employs time evolution along a closed
contour, so one has to keep two copies of each field: one for the
forward and another for the backward evolution. In a classical context
the Martin--Siggia--Rose
\cite{MSR,Janssen76,DeDominics,Janssen79,BJW} method requires an
additional set of fields to resolve the functional $\delta$--functions
of Langevin equations. Most importantly for the present subject, in
reaction--diffusion kinetics the Doi--Peliti \cite{Doi76,Peliti84}
operator technique deals with the creation and annihilation
operators for each reagent. Thus it employs two variables (or one
complex field) for every {\em real} physical degree of freedom. (For
a discussion of the connections between these techniques see, e.g.,
Refs.~[\onlinecite{Kamenev00,Kamenev05}].)

An important observation is that in all these examples the two
sets of fields (being properly transformed) may be considered as
canonically conjugated variables. As a result, instead of the
equilibrium order parameter $q(x)$, one has to deal with the
canonical pair: $q(x,t)$ and $p(x,t)$. Correspondingly a
reaction--diffusion system may be described by the {\em
Hamiltonian} action
\begin{equation}\label{nonequilibrium}
  S=\int \!dt\! \int\!\! d^d x \left[\,p\,\partial_t q +D\nabla p\nabla q - H_R(p,q)\,\right]\, ,
\end{equation}
where the Hamiltonian $H_R(p,q)$ is determined by the set of
reactions specific for a given model (see below).

Comparing Eqs.~(\ref{nonequilibrium}) and (\ref{equilibrium}), one
notices that the reaction Hamiltonian $H_R(p,q)$ plays a role
similar to the effective potential $V(q)$ in equilibrium
statistical mechanics. Thus it is plausible that $H_R(p,q)$ may
encode information about possible nonequilibrium transitions
in a way analogous to what $V(q)$ does. Specifically, one wants to
know what is the Hamiltonian analog  of the potential minima,
given by $\partial_q V=0$,  in the Lagrangian formulation. The
answer is that it is the {\em classical} equations of motion:
$\partial_t q =
\partial_p H_R$ and $\partial_t p=-\partial_q H_R$. One is
looking, therefore, for a geometric way to picture the Hamiltonian
equations of motion. We argue below that the way to do it (at
least for the one--component models) is to consider the phase
space trajectories in the $(p,q)$ plane. Indeed, the classical
equations of motion conserve ``energy." Thus the phase--space
trajectories are given by the curves $H_R(p,q)=\mbox{const}$.
Moreover, for systems with absorbing states the only trajectories
which may intersect correspond to  zero energy. As a result, the
set of curves
\begin{equation}\label{H=0}
  H_R(p,q)=0
\end{equation}
determines entirely the topology of the phase space.

The  main message  of this paper is that the curves specified by
Eq.~(\ref{H=0})  and the corresponding topology of the phase
portrait classify possible phase transitions in
reaction--diffusion models. It is the  web of the zero ``energy''
trajectories which plays the role of  minima of the $V(q)$ potential
in the equilibrium statistical mechanics. A topological
rearrangement of this web as a function of the control parameter
signals the existence of a phase transition. The corresponding
topology  is in one-to-one correspondence with the universality
classes.

Below we show that the number of distinct generic phase--space
topologies (for one-component systems) is rather limited, indicating
that all possible universality classes may be exhausted. Some of the
topologies correspond to first--order transitions, while others
to continuous ones [much the way potential (\ref{potential})
contained both]. The latter class may develop nontrivial critical
exponents below a certain critical dimension $d_c$. The way to find
these exponents is to follow the RG flow of  constants of the
Hamiltonian $H_R(p,q)$ upon elimination of the small--scale
fluctuations. While the Hamiltonian itself may be complicated, it is
only the topology of the phase space that matters, not the specific
shape of the curves. Any given topology may be modeled by a simple
polynomial of $q$ and $p$, again much the way the simple polynomial
(\ref{potential}) suffices to describe many equilibrium systems.
Thus one must follow only  changes of the topology of the phase
portrait upon RG transformations. One should also verify that a
given topology is stable upon RG transformations, i.e., that it cannot be
reduced to a more generic one by decimation. The resulting 
fixed--point topologies and corresponding fixed--point Hamiltonians
$H^*_R(q,p)$ provide information about the universality
classes.

We found that the DP universality class  (represented by the
simplest triangular structure on the phase plane) serves as
parent for a family of  descending classes. Each subsequent class
in the family is characterized by a  minimal number $k$ of
particles needed to initiate reactions. We denote it as the
$k$-particle contact process with diffusion (kCPD). Here 1CPD is
DP, while 2CPD is the PCPD
(for a   review see Ref.~[\onlinecite{HH04}] and references
therein). According to scaling analysis, above the upper critical
dimension $d_c=4/k$ the kCPD's are characterized by
mean--field critical exponents, e.g., $\beta = 1$ and $\nu_\bot =
k/2$. For $d\leq d_c$ and $k=2,3$ we found that RG flows to a
strong--coupling fixed--point that cannot be accessed in the
$\epsilon$ expansion (see also Refs.~[\onlinecite{HH04,JvWOT04}]).
We also discuss the possible nature of the strong-coupling fixed
point for $k=2$.

Similarly the PC universality class generates
a family of  classes, characterized by  a minimal number $k$, of
incoming particles required  for {\em all} reactions. We call them
$k$-particle parity conserving (kPC). Their upper critical
dimension is $d_c=2/k$. In addition to kCPD and kPC we identify
two more families of universality classes. They both originate
from  reversible reactions which may go in both directions with
different rates. We call them $k$-particle reversible (kR) and
$k$-particle reversible parity conserving (kRPC). In both cases
$k$ stays for a minimal number of incoming particles. Their critical
dimensions are $2/k$ and $2/(k+1)$ correspondingly.

These four families seem to exhaust all possible continuous
transitions reachable by the tuning of a {\em single} control
parameter and capable of exhibiting a {\em non--mean--field}
behavior. This means that any phase portrait, topologically
different from that of the four families, is unstable upon
renormalization. In the large--scale limit it flows towards one of
the  stable topologies. The latter are protected by  certain
symmetries of the action against deformations introduced by the
RG. The fluctuation-induced renormalization may be not effective
if the space dimensionality is sufficiently high. As a result,
other topologies may appear to be stable and lead to different
universality classes. However, by virtue of the ineffectiveness of
fluctuations, such universality classes are bound to exhibit
trivial {\em mean--field} behavior.

The paper is organized as follows:  in Sec.~\ref{section_PP} we
introduce reaction Hamiltonians and their phase portraits. Section
\ref{section_FOT} is devoted to models exhibiting the first--order
transitions and discusses the topological structure of their phase
portraits. In Sec.~\ref{section_DP} models of the DP
universality class and its derivatives, kCPD's, are considered. We
demonstrate that the triangular topology of the phase portrait is the
typical feature of all universality classes of this kind. In
Sec.~\ref{section_PC} we consider the parity conserving model and
its generalizations, kPC's. The rectangular phase portrait topology
of the reversible reaction models, kR's and kRPC's, is discussed
in Sec.~\ref{sec_rev}. Finally some conclusions and outlook are
drawn in Sec.~\ref{section_conclusion}.

\section{Reaction Hamiltonians and phase portraits}
\label{section_PP}

The standard way to introduce the ``quantum'' reaction Hamiltonian
is by employing the creation and annihilation operator technique of
Doi and Peliti \cite{Doi76,Peliti84,LeeCardy,Cardy97,Mattis98}.
Here we choose to follow a different, though completely
equivalent, strategy \cite{ElgartKamenev04}. Consider a generic
reaction that transforms $k$ particles into $m$ equivalent ones
with the probability $\lambda$:
\begin{equation}
    k A \stackrel{\lambda}{\rightarrow} m A,\label{Reaction}
\end{equation}
The corresponding master equation for the temporal evolution of
the probability  ${\cal P}_n(t)$ of a configuration with $n$ particles
has the form
\begin{equation}
    \partial_t {\cal P}_n(t) =
    \lambda \left[{n+k-m\choose k} {\cal P}_{n + k - m}(t) -
    {n\choose k} {\cal P}_n(t)\right]\, .
                                             \label{master_gen}
\end{equation}
The two terms on the right hand side (RHS) represent the probabilities of ``in''
and ``out'' processes correspondingly. The master
equation~(\ref{master_gen}) is to be supplemented with an initial
distribution, e.g., ${\cal P}_n(0) = e^{-n_0}{n_0^n}/{n!}$,  the Poisson
distribution  with the mean value $n_0$, or
${\cal P}_n(0)=\delta_{n,n_0}$,  the fixed initial particle number.

Let us define now a {\em generating function} as
\begin{equation}
\label{generating}
    G(p,t) \equiv \sum\limits_{n=0}^\infty p^n {\cal P}_n(t)\, .
\end{equation}
The parameter $p$ will play the role of the canonical momentum; so far
it has been introduced pure formally. The value $p=1$ plays a special
role. First, the conservation of probability implies the
normalization condition
\begin{equation}
    G(1,t)\equiv 1\, .
                                                     \label{norm}
\end{equation}
Second, the moments of ${\cal P}_n(t)$ may be expressed through
derivatives of the generating function at $p=1$, e.g., 
\begin{equation}
    \langle n(t)\rangle \equiv\sum_n n {\cal P}_n(t) = 
    \partial_p G(p,t)|_{p=1}\,.
\end{equation}
Knowing the generating function, one may find a
probability of having (integer) $n$ particles at time $t$ as
${\cal P}_n(t)=
\partial_p^n G(p,t)|_{p=0}/n!$.

In terms of the generating function the  master
equation~(\ref{master_gen}) may be {\em identically} rewritten as
\begin{eqnarray}
    \partial_t G =  \hat{H}_R(p,\hat q)\, G\,,\label{Ham}
\end{eqnarray}
where the non--Hermitian normally ordered operator $\hat{H}_R$
stays for
\begin{eqnarray}
   \hat{H}_R(p,\hat q)  =  \frac{\lambda}{k!}
    (p^m-p^{k})\,\hat{q}^k\,.\label{Gen_Fun}
\end{eqnarray}
Here we have introduced the ``coordinate'' operator
\begin{equation}
    \hat{q} \equiv \frac{\partial}{\partial p}\, ,
\end{equation}
obeying  the canonical commutation relation $[\hat q,p]=1$.
Because of the obvious analogy with the Schr\"odinger  equation,
we shall refer to the operator $\hat{H}_R$ as the Hamilton
operator in the $p$ representation. From the normalization
condition, Eq.~(\ref{norm}), it follows that
\begin{equation}\label{norm2}
    \left.\hat{H}_R(p,\hat q)\right|_{p=1} = 0\,.
\end{equation}
Any Hamiltonian derived from a probability conserving master
equation  necessarily satisfies  this property.

One can easily generalize this construction for the case where
many reactions with rates $\lambda_{k m}$ take place at the
same time. To this end one can simply algebraically add the
corresponding partial Hamiltonians to obtain the full reaction
Hamiltonian. If there is no particle production from the vacuum, i.e.,
$k\neq 0$ for any $m$--the empty state with $n=0$ is an
absorbing state in the sense that the system can never leave it.
According to Eq.~(\ref{Gen_Fun}) any Hamiltonian function
describing a system with empty absorbing state must satisfy
\begin{equation}\label{norm3}
    \left.{H}_R(p, q)\right|_{q=0} = 0\,
\end{equation}
in addition to Eq.~(\ref{norm2}).

Before  considering the full ``quantum'' problem, Eq.~(\ref{Ham}),
let us analyze  the corresponding classical dynamics. The
classical equations of the motion are
\begin{eqnarray}
    \partial_t{q} &=& \,\,\frac{\partial}{\partial p}\,H_R(p,q)\,,\label{motion_q}\\
    \partial_t{p} &=& \!\!- \frac{\partial}{\partial q}\, H_R(p,q)\,.\label{motion_p}
\end{eqnarray}
Due to  Eqs.~(\ref{Gen_Fun}) and (\ref{norm2}), $p=1$ is always
one of the solutions of Eq.~(\ref{motion_p}). Substituting $p=1$
into Eq.~(\ref{motion_q}), one obtains [for the Hamiltonian~(\ref{Gen_Fun})]
\begin{equation}
    \partial_t{q} =  \frac{\lambda}{k!}\,(m-k)  \, q^k\, .
 \label{mfrate}
\end{equation}
This is nothing but the mean--field rate equation for the average
particle number $\langle n(t)\rangle$, neglecting all fluctuation
effects. Therefore one may identify the variable $q$ as the
reaction ``coordinate'' (in fact this notation is not  precise,
since it is true only along the line $p=1$).

\begin{figure}
\includegraphics[width=6cm]{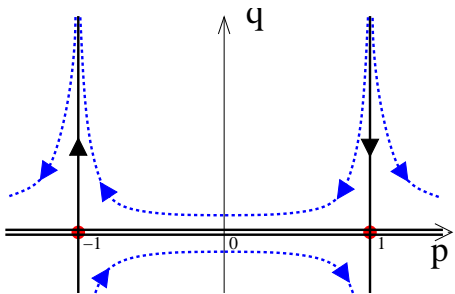}
\caption{(Color online) Phase portrait of the binary annihilation
system, $2 A \stackrel{\lambda}{\rightarrow} \emptyset$. The
corresponding classical  Hamiltonian is given by  $H_R(p,q) =
\frac{\lambda}{2}(1-p^2)q^2$. Solid black lines show zero--energy
trajectories: generic lines $p= 1$ and double degenerate $q=0$ and
the ``accidental'' line $p=-1$. Dashed  colored curves indicate
trajectories with  nonzero energy. The arrows show the evolution
direction.} \label{fig_ann}
\end{figure}

In order to proceed with the classical problem,
Eqs.~(\ref{motion_q}) and (\ref{motion_p}), beyond the reaction rate
approximation ($p=1$),  it is convenient to consider the phase
space of the system. The classical equations of motion
(\ref{motion_q}) and (\ref{motion_p}) conserve energy. As a
result, the  phase--space $(p,q)$ evolution of the system takes
place along the trajectories given by $H_R(p,q) = \mbox{const}$,
where the constant energy is determined from the initial
conditions. Among all possible trajectories the ones with $H_R=0$
play a special role. For one thing, the evolution prescribed by
the rate equation ($p=1$) takes place along one of such lines
[cf. Eq.~(\ref{norm2})]. More importantly, the trajectories with
$H_R=0$ may intersect each other. Indeed, the two zero-energy
lines, guaranteed by Eqs.~(\ref{norm2}) and (\ref{norm3}), i.e.,
$p=1$ and $q=0$--intersect in the point $(1,0)$. Therefore the set
of intersecting zero-energy curves plays the role of separatrix;
i.e., it divides the entire phase space on the isolated sectors.
All other trajectories cannot intersect the zero-energy ones and
are confined to one of the sectors. The web of the zero-energy
curves uniquely determines the topology of the phase portrait. An
example of such a construction is given in Fig.~\ref{fig_ann}.

Going back to the full ``quantum'' problem,  one may write a
formal solution of Eq.~(\ref{Ham}) as
\begin{equation}\label{green}
    G(p_f,t_f) = \int U(p_f,t_f;p_0,t_0)\,G_0(p_0)\,
    \mathrm{d}p_0\,,
\end{equation}
where the Green function $U(p_f,t_f;p_0,t_0) $ is given by the
$T$ exponent: $T\exp\{\hat H_Rt\}$. Dividing the time interval
$[t_0,t_f]$ into $N\to \infty$ steps and introducing the resolution
of unity at each one of them, one obtains the Feynman
representation
\begin{equation}\label{funintegral}
    U(p_f,t_f;p_0,t_0) = \int\! \mathcal{D} p(t)\, \mathcal{D} q(t) \,\, e^{\,-S[p,q]}\ ,
\end{equation}
with the Hamiltonian action
\begin{equation}\label{action0d}
    S[p,q] = \int_{t_0}^{t_f} \! dt \left[ p\, \partial_t{q} - H_R(p,q) \right]\,.
\end{equation}

To combine reaction kinetics with the random walk on a lattice,
one needs to modify the master equation. The corresponding
generating function becomes a function of many variables $p_i$,
where the index $i$ enumerates the lattice sites. One may also
introduce the conjugated variables $\hat q_i=\partial/\partial
p_i$. The resulting Hamiltonian takes the form
\begin{equation}
    \hat H = -\tilde D\sum\limits_{\langle i,j\rangle} (p_i - p_{j})(\hat q_i - \hat q_{j})+
    \sum\limits_i \hat H_R(p_i,\hat q_i) \,,
\end{equation}
where $\tilde D$ is a hopping probability per unit time and the
sum in the first term on the RHS runs over nearest neighbors
$i,j$.
Taking the continuum  limit \cite{Doi76,Peliti84} and introducing
the pair of canonically conjugated fields $p(x,t)$ and $q(x,t)$,
one arrives at the quantum field theory with the Hamiltonian
action (\ref{nonequilibrium}). The diffusion constant in
Eq.~(\ref{nonequilibrium}) is given by $D=\tilde Da^2$, where $a$
is the lattice constant.

Unless the system is very close to extinction, the functional
integral in Eq.~(\ref{funintegral}) may be evaluated in the saddle
point approximation. In such a case the Green function is given by
the exponentiated action of a classical trajectory, satisfying the
proper boundary conditions \cite{ElgartKamenev04} [much the same
way as the minima of the potential $V(q)$, Eq.~(\ref{potential}),
dominate the partition function away from an equilibrium phase
transition]. A possible phase transition may be associated with a
qualitative  change in the behavior of the phase-space
trajectories (cf. the rearrangement of minima of the
potential upon an equilibrium transition). In other words, phase
transitions lead to a change  of topology of the phase--space
portrait. Since the latter is determined by the set of the
zero-energy lines, it is the rearrangement of this set, upon
variation of a control parameter, which must be associated with
the phase transition.

If a system is close enough to a phase transition (or extinction),
the saddle point approximation may lose its validity (below
critical dimensionality $d_c$). One can then employ  the RG technique
to progressively integrate out the small scale fluctuations. Upon
such a procedure the constants and even the functional form of the
reaction Hamiltonian flow. However, one needs to follow the
topology of the phase space, rather than a specific form of the
trajectories. Around the transition the topology may be fully
encoded in a relatively simple polynomial, which in turn provides
a full characterization of the transition (at least for small
$\epsilon=d_c-d$). Considering all distinct topologies, stable
upon RG transformations, one may classify the possible phase
transitions.

We turn now to an illustration of these ideas with specific examples.

\section{Models with  First--Order Transitions}\label{section_FOT}

\begin{figure}
\includegraphics[width=8cm]{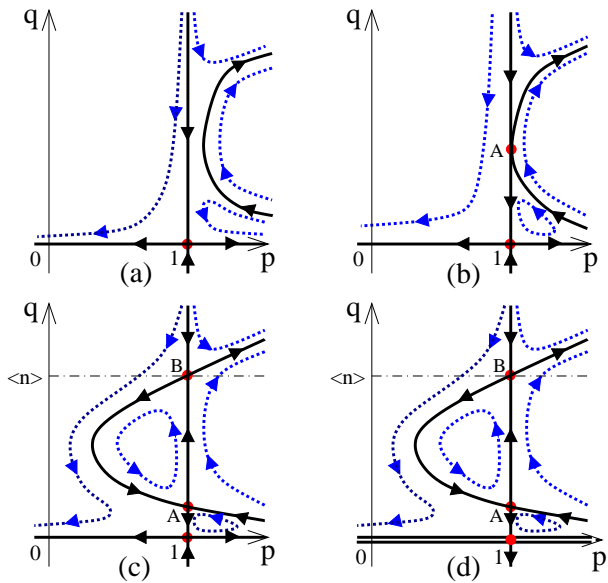}
\caption{(Color online) Phase portraits of the system
 exhibiting the first--order
phase transition. Thick solid lines represent the zero-energy
trajectories. (a)-(c) the set, Eq.~(\ref{first_order_II_example}):
(a) extinction phase, (b) transition point, and (c) active phase. (d)
Reaction of the type of Eq.~(\ref{first_order_II}) with $k=2$,
$n=3$, and $i=4$ in the active phase.  }
 \label{fig_fot_II}
\end{figure}

Consider a set of reactions, given by
\begin{equation}\label{first_order_II_example}
  A \stackrel{\lambda}{\rightarrow} \emptyset  \,,\quad \,\,
  2 A \stackrel{\mu}{\rightarrow} 3 A  \,,\quad \,\,
  3 A \stackrel{\sigma}{\rightarrow} 2 A  \,.
\end{equation}
The corresponding  reaction Hamiltonian, according to
Eq.~(\ref{Gen_Fun}), may be written as
\begin{equation}
    H_{R} = \left(\lambda - {\mu\over 2}\, p^2 q +
    {\sigma\over 6}\, p^2 q^2\right)(1-p)\,q\,.
\end{equation}
There are three  zero--energy curves: the two generic lines $p=1$
and $q=0$, following from Eqs.~(\ref{norm2}) and (\ref{norm3}),
and the additional ``accidental'' curve $p^2=2\lambda/(\mu
q-\sigma q^2/3)$. The phase portrait of the system is depicted in
Figs.~\ref{fig_fot_II}(a)-~\ref{fig_fot_II}(c). Its topology is qualitatively
different for $\mu < \mu_c=\sqrt{8\lambda\sigma/3}$ and $\mu >
\mu_c$. In the former case, Fig.~\ref{fig_fot_II}(a), the
``accidental'' curve does not intersect the mean-field line $p=1$
and the flow along the latter is directed towards $q=0$
(extinction). For $\mu=\mu_c$ the two curves touch each other,
Fig.~\ref{fig_fot_II}(b), creating a stationary point with a
finite concentration $q=3\mu/(2\sigma)$. At a larger creation rate
$\mu>\mu_c$, there is a stable point with concentration
$\langle n\rangle$, Fig.~\ref{fig_fot_II}(c).

This is the first--order transition scenario. Indeed,  the stable
concentration experiences a discontinuous jump from zero at
$\mu<\mu_c$ to some finite value at $\mu\geq\mu_c$. It may be
generalized to any reaction set of the form
\begin{equation}\label{first_order_II}
  k A {\rightarrow} (k-l) A  \,,\quad \,\,
  n A {\rightarrow} (n+m) A  \,,\quad \,\,
  i A {\rightarrow} j A  \,,
\end{equation}
where the first two reactions  satisfy the  condition $k<n$ and
the last reaction is necessary to ensure a finite particle density
by having $i>n,j$ (for hard--core ``fermionic'' models this
restriction is intrinsic and the last reaction  may be omitted).
The corresponding reaction Hamiltonian  may be written as
\begin{equation}\label{hamiltonian_fot_II_gen}
    H_{R} = \left[h_1(p) - h_2(p)q^{n-k}+
    h_3(p)q^{i-k}\right](1-p)\,q^k\,,
\end{equation}
where the functions $h_1,h_2$, and $h_3$ are positive in the interval $p \in
[0,1]$. An example of the phase portrait in the active phase and
$k=2$ is depicted in  Fig.~\ref{fig_fot_II}(d).

Strictly speaking, the condition $\mu=\mu_c$
[Fig.~\ref{fig_fot_II}(b)] signifies the appearance of the {\em
metastable} state (unless $D\to \infty$). Such a metastable
state becomes stable at some larger creation rate $\tilde
\mu_c(D)>\mu_c$. Therefore, in terms of the bare system's
parameters the actual first--order transition takes place when the
phase portrait has a form of Fig.~\ref{fig_fot_II}(c).
Alternatively, one may imagine integrating out spatial
fluctuations (governed by the diffusion constant $D$) and plotting
the phase portrait of the effective zero--dimensional ($d=0$) system in terms of the {\em
renormalized} parameters. It is in this latter sense that the
transition point is depicted by Fig.~\ref{fig_fot_II}(b).

\begin{figure}
\includegraphics[width=8cm]{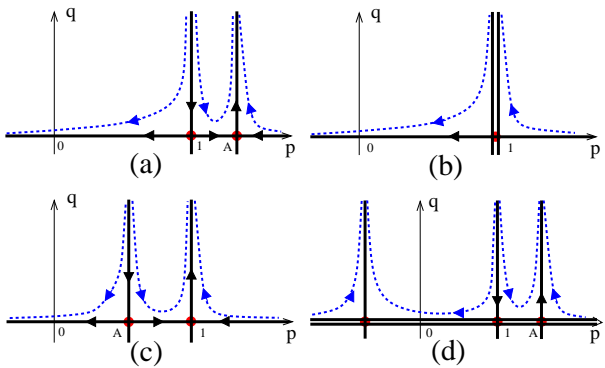}
\caption{(Color online) Phase portraits of the system,
Eq.~(\ref{firstordergeneralized}), exhibiting the first--order
phase transition. Thick solid lines represent the zero-energy
trajectories. (a)--(c) Reaction set with $k=l=m=1$: (a) extinction
phase, (b) transition point, and (c) unlimited proliferation phase.
(d) Reaction set with  $k=l=2$ and $m=1$ in the extinction phase.}
 \label{fig_fot}
\end{figure}

An interesting limiting case \cite{Paes03b,BSH06} of the first--order
transition is the reaction set
\begin{equation}\label{firstordergeneralized}
  k A \stackrel{\lambda}{\rightarrow} (k-l) A  \,,\quad \,\,
  k A \stackrel{\mu}{\rightarrow} (k+m) A  \,,
\end{equation}
described by the Hamiltonian
\begin{equation}\label{hamiltonian1gen}
    H_{R} = h(p)(1-p)\, q^k \, ,
\end{equation}
where $h(p)$ is a polynomial the degree $k+m-1$. The zero-energy
lines are given by $p=1$, the $k$ times degenerate line $q=0$
along with the lines $p=p_i$, where $p_i$ are  roots of the
polynomial $h(p_i)=0$, Fig.~\ref{fig_fot}. It is easy to check
that for
 $   \mu_c = \frac{l}{m}\lambda\, $
one of the roots of the polynomial is $p_1=1$. In this situation
the $p=1$ zero-energy line is doubly degenerate, corresponding to
the first--order transition. Upon such a transition the $p=1$ and
$p=p_1$ lines interchange their relative positions and the
extinction behavior turns into the unlimited proliferation one.

Notice that adding a proliferation-restricting reaction
$iA\rightarrow jA$, with $i>k$, to the set
(\ref{firstordergeneralized}) qualitatively changes its phase
portrait. In general, such a modification leads to a continuous
phase transition [see, e.g., Eqs.~(\ref{DP}) and (\ref{PCPD}) below]. An
analogous transformation of the first--order to the continuous 
transition takes place for restricted (``fermionic'') reaction rules
\cite{Racz02,Odor03}. This is consistent with the conjecture that
the ``fermionic'' condition implies an effective annihilation process
of the order of $k+1$; see, e.g.,
Refs.~[\onlinecite{Hinrichsen00,Odor04}] and references therein.

The phase portrait description given above has the status of the
mean-field consideration only (much the same way as the finding of
minima of the effective potential, Eq.~(\ref{potential}), in the
equilibrium theory). It does {\em not} exclude the possibility that
below some lower  critical dimension $d_c'$ the fluctuations may
drastically change the system's behavior. In particular, a
transition, being first--order in the mean-field scenario,  may be
turned to a continuous one by fluctuation effects. This may
explain the apparent continuous transition seen in $d=1$ systems,
similar to those given by Eq.~(\ref{first_order_II})
\cite{Kock02,Odor_pr,Dornic_pr}. It was also observed in some
fermionic reaction schemes that the first--order transition
observed at large $D$ turns out to be continuous for $D<D_c$
\cite{Tome91,Cardozo06}. The subject definitely needs more
investigation both analytical and numerical.

\section{Directed Percolation and its generalizations}\label{section_DP}

\subsection{DP models}

Consider a reaction set which includes death, branching, and
coalescence reactions:
\begin{equation}
    A \stackrel{\lambda}{\rightarrow} \emptyset\,, \quad\,\,
    A \stackrel{\mu}{\rightarrow} 2 A\,, \label{DP}\quad\,\,
    2 A \stackrel{\sigma}{\rightarrow} A\,.
\end{equation}
The corresponding reaction Hamiltonian takes the form
\begin{eqnarray}\label{hamiltonianDP}
    {H}_{R} &=& \lambda (1-p)\,{q} + \mu (p^2-p)\,{q}+ {\sigma\over 2} (p-p^2)\,{q}^2\nonumber \\
    &=&   \left(\lambda -\mu p +{\sigma\over 2}\, p\,q \right)(1-p)\,q\, .
\end{eqnarray}

\begin{figure}
\includegraphics[width=8cm]{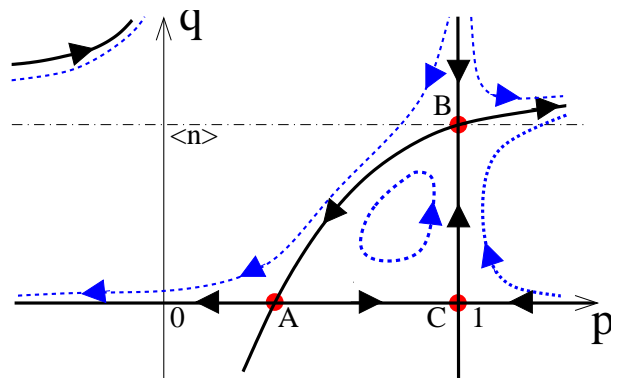}
\caption{(Color online) Phase portrait of the DP system in the
active phase. Thick solid lines represent zero-energy
 trajectories which divide the phase space into a number of disconnected regions.
 Point $B=\{1, 2(\mu-\lambda)/\sigma\}$ represents the active mean-field point.
 The system is brought to the phase transitions if points $A$, $B$, and $C$ coalesce.  }
 \label{fig_dp}
\end{figure}

The phase portrait of the DP system is depicted in
Fig.~\ref{fig_dp}. The lines of zero energy are generic $p=1$ and
$q=0$ trajectories along with the ``accidental'' trajectory
$q=2(\mu p-\lambda)/\sigma p\,$. According to the mean-field analysis
[classical equations~(\ref{motion_q}) and (\ref{motion_p}) with
$p=1$], there is an active phase with the average density
\begin{equation}
                                \label{av_density}
\langle n \rangle =2\,{\mu-\lambda \over \sigma}\,
\end{equation}
 for $\mu > \lambda$. The
active phase corresponds to point $B$ in Fig.~\ref{fig_dp}. The
system may be brought to extinction by driving  the control
parameter $m=\mu-\lambda$ through zero. Therefore the
$\mu=\lambda$ point corresponds to the continuous phase
transition. The transition  is represented by the phase portrait
with the three zero-energy trajectories intersecting at the {\em
single} point $(1,0)$. According to Eq.~(\ref{av_density}) the
mean-field order-parameter exponent is $\beta=1$. The other
mean-field exponents \cite{Hinrichsen00} are $\nu_\perp = 1/2,\
\nu_\| = 1$.

To go beyond the mean--field picture one needs to investigate the
immediate vicinity of the phase transition. To focus on this
regime it is convenient to shift the momentum variable
\begin{equation}\label{shift}
  p-1 \rightarrow p\, .
\end{equation}
Moreover, close to the transition the phase portrait and thus the
Hamiltonian may be modeled by the three  intersecting {\em
straight} lines, Fig.~\ref{fig_dp_gen}(a)-~\ref{fig_dp_gen}(c). This way one arrives
at the model Hamiltonian, applicable close to the DP transition:
\begin{equation}\label{genericDP}
    H_R = (m+up-vq)\, p\,  q \, .
\end{equation}
The bare values of the constants are given by $m=\mu-\lambda$,
$u=\mu$, and $v=\sigma/2$. The corresponding action,
Eq.~(\ref{nonequilibrium}), acquires the form {\cite{Hinrichsen00,THL05,JanssenTauber05}
\begin{equation}\label{Reggeon}
    S = \!\int \!\!dt\, d^d x\left[p(\partial_t - D\nabla^2)\,q - m p\, q  - u p^2\, q + v p\,
    q^2\right],
\end{equation}
which may be recognized as a Reggeon field theory action
\cite{CardySugar80,Grassberger97}.

There are many other reaction sets, in addition to Eq.~(\ref{DP}),
with the same ``triangular'' topology of the phase portrait.  Some
of the examples are $    A\to \emptyset\,;\quad A\to
(m+1)A\,;\quad 2A\to A\,,\quad m > 1\,, $ or $  A\to 2 A\,;\quad
2A\to \emptyset\,$ (see also Sec.~\ref{section_critpoint});
etc. In the vicinity of the phase transition they all exhibit the
topology of the phase portrait depicted in
Figs.~\ref{fig_dp_gen}(a)-~\ref{fig_dp_gen}(c). Therefore they all may be described by
the model Hamiltonian~(\ref{genericDP}). Accordingly they all
belong to  the same DP universality class.

Naive scaling dimensions of the action (\ref{Reggeon}) are $z=2$,
while $[p]+[q]=d$. Since one expects \cite{Odor04a} both vertices
$u$ and $v$  to have the same scaling dimensions, one finds
$[p]=[q]=d/2$. As a result, the bare scaling dimensions of the
vertices are $[m]=2$, while $[u]=[v]=2-d/2$. Therefore below the
critical dimension $d_c=4$ the nonlinear vertices $u$ and $v$ are
relevant and the mean--field treatment is expected to fail.

\begin{figure}
\includegraphics[width=8cm]{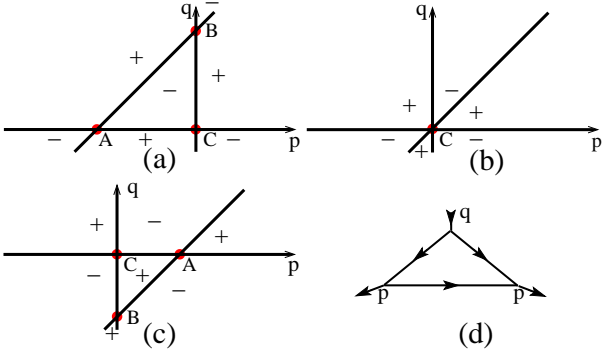}
\caption{Generic phase portrait of DP models in the vicinity of
the phase transition [after the shift, Eq.~(\ref{shift})]. (a)
Active phase, $m>0$; (b) transition point, $m=0$; (c) extinction
phase, $m<0$. The plus and minus signs show the sign of the
Hamiltonian  in each sector. (d) The one--loop diagram
renormalizing $u$ vertex (vertices $m$ and $v$ are renormalized in
a similar way). }
 \label{fig_dp_gen}
\end{figure}

The one--loop corrections to the naive scaling are given by the
triangular diagrams, like the one depicted in
Fig.~\ref{fig_dp_gen}(d). Such diagrams are logarithmically
divergent in $d=4$, as expected. Straightforward calculations
\cite{Hinrichsen00,THL05} (see Ref.~[\onlinecite{J81}] for the
two--loop approximation) lead to the following set of RG
equations:
\begin{eqnarray}\label{RGmuv}
  \partial_l m &=& (2 - S u v)\,m\, , \label{m} \\
  \partial_l u &=& (\epsilon/2 - 2 S uv)\,u\, ,\label{u}\\
  \partial_l v &=& (\epsilon/2 - 2 S uv)\,v\, , \label{v}
\end{eqnarray}
where $\epsilon=4-d$ and the differentiation is over the logarithm
of the scaling factor. We have introduced the factor
\begin{eqnarray}\label{Sdp}
     S = \partial_l\frac{1}{4}\int_{\Lambda e^{-l}}^\Lambda \frac{\mathrm{d}^d k}{D^2 k^4}\,
      = \ \frac{\Lambda^{d-4}}{32 \pi^2 D^2}\,,
\end{eqnarray}
which may be absorbed in the proper redefinition of the running
constants.

According to Eqs.~(\ref{u}) and (\ref{v}), $\partial_l(u/v)=0$, meaning
that the slop of the ``accidental''  zero-energy line, $q=(m+up)/v$,
remains intact upon the renormalization procedure. In fact,  this
statement is exact  because of the symmetry \cite{foot1}. As a
result, the overall topology of the phase portrait is preserved by
the RG. For $d<4$ the RG equations (\ref{m})--(\ref{v})
predict a nontrivial fixed point given by $m^*=0$, $u^* =
\sqrt{\epsilon\mu/(4 S\sigma)}$, and $v^* = \sqrt{\epsilon\sigma/(4
S\mu)}$. Substituting these values into Eq.~(\ref{genericDP}), one
finds the fixed--point reaction Hamiltonian $H^*_R(p,q)$
corresponding to the DP universality class phase transitions. Its
phase portrait is depicted in Fig.~\ref{fig_dp_gen}(b). Linearizing
the RG equations~(\ref{m})--(\ref{v}) near the fixed point,  one
finds $\partial_l m = \nu_\perp^{-1} m$ with the critical exponent
$\nu_\perp = (2 - \epsilon/4)^{-1}\approx 1/2+\epsilon/16$. The
other critical exponents may be deduced in the standard way
\cite{Hinrichsen00,THL05}, e.g., $\beta \approx 1 - \epsilon/6$.

\subsection{$k$-particle contact  processes}\label{seckcpd}

As  mentioned in the Introduction, the DP universality class is
extremely robust. This fact is due to the stability of the
``triangular'' topology of the  phase portrait near the
transition. One may try to change this topology by, say, requiring
four or more zero-energy trajectories to intersect in the same
point. It is clear, however, that in general one must fine-tune
more than one parameter to reach such a scenario. Even if bare
reaction rates are specially chosen to let it happen, the tuning
is not expected to survive upon RG integration of  fluctuations.
Therefore crossing of more than three lines is possible only at a
multicritical transition point.  The only way to go beyond the DP
is if a different topology of the phase portrait is imposed by an
additional {\em symmetry}.

In this section we discuss a class of models where the {\em
minimum} number of particles needed to initiate {\em any} reaction
is restricted to be $k>1$. According to Eq.~(\ref{Gen_Fun}) {\em
all} $\,$ terms in the corresponding reaction Hamiltonian must
contain $\hat q^k$ or higher power. In terms of the phase portrait
it means that the generic $q=0$ zero-energy trajectory is
$k$ times degenerate. To emphasize the difference with
nondegenerate lines, we  denote such $k$-degenerate trajectories
by $k$ closely spaced parallel lines. An important fact is that
the degeneracy is preserved by the RG transformations. Indeed, the
fluctuations cannot initiate a reaction with fewer than $k$
incoming particles, if it is not in the original reaction set. We
denote such models as $k$-particle contact processes with diffusion
(kCPD).

 \begin{figure}
\includegraphics[width=8cm]{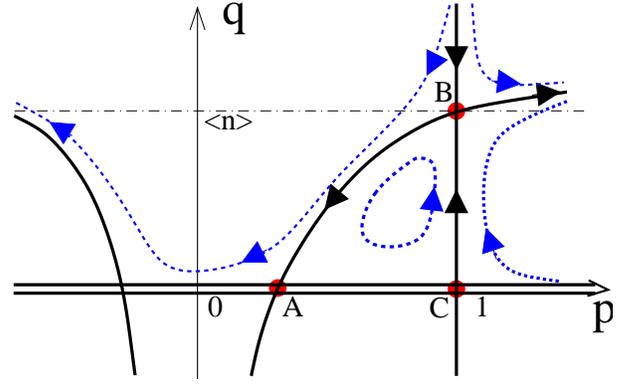}
\caption{(Color online) Phase portrait of the 2CPD system in the
active phase (cf. DP, Fig.~\ref{fig_dp}). The zero-energy line
$q = 0$ is doubly degenerate and is depicted by the double line. At
the transition points $A$, $B$, and $C$ coalesce. }\label{fig_pcpd0}
\end{figure}

To be specific, let us consider the case of $k=2$, which is
represented by, e.g., the following set of reactions:
\begin{equation}
    2 A \stackrel{\lambda}{\rightarrow} \emptyset\,; \quad\,\,
    2 A \stackrel{\mu}{\rightarrow} 3 A\, ;\label{PCPD}\quad\,\,
    3 A \stackrel{\sigma}{\rightarrow} 2 A\, .
\end{equation}
The corresponding reaction Hamiltonian takes the form:
\begin{eqnarray}\label{hamiltPCPD}
    {H}_{R} &=& {\lambda\over 2} (1-p^2)\,{q^2} + {\mu\over 2} (p^3-p^2)\,{q^2}
    + {\sigma\over 6} (p^2-p^3)\,{q}^3\nonumber \\
    &=&  {1\over 2} \left(\lambda\,(1+p) - \mu\, p^2 +{\sigma\over 3}\, p^2\,q \right)(1-p)\,q^2\, .
\end{eqnarray}

The phase portrait of the 2CPD system is depicted in
Fig.~\ref{fig_pcpd0}. The lines of zero energy are generic $p=1$
and {\em double--degenerate} $q=0$  trajectories.   The
``accidental'' trajectory is given by $q=3\left[\mu
p^2-\lambda(1+p)\right]/(\sigma p^2)\,$. There is an active phase with
average density $ \langle n \rangle =(3\mu-6\lambda)/ \sigma$
for $\mu > 2\lambda$. It corresponds to point $B$ in
Fig.~\ref{fig_pcpd0}. The system may be driven to extinction
by tuning  the control parameter $m=\mu/2-\lambda$ to zero.
Therefore the $\mu/2=\lambda$ point corresponds to the continuous
phase transition. At the transition point the {\em four} (we count
the $q=0$ line twice) zero-energy lines are intersecting in the single
point $(1,0)$.

Focusing on the transition region, one may shift the momentum
variable $p-1 \rightarrow p$ and model the zero-energy
trajectories by straight lines, Fig.~\ref{fig_kcpd}(a). The
resulting model Hamiltonian, applicable close to the  transition,
is
\begin{equation}\label{PCPDham}
    H_R = (m+up-vq)\, p\,  q^2 \, .
\end{equation}
The bare values of the constants are given by $m=\mu/2-\lambda$,
$u=\mu-\lambda/2$, and $v=\sigma/6$. Since near the transition
$m\approx 0$ and thus $\mu\approx 2\lambda >0$, one finds for the
bare value $u\approx 3\lambda/2>0$. Apart from Eq.~(\ref{PCPD}),
there are infinitely many other reaction sets, which have the same
topology of the phase portraits.  Therefore  the phase transition
of these other models  is described by the same model
Hamiltonian~(\ref{PCPDham}). Examples include $2A\to A\,,\, 2A\to
4A\,,\, 4A\to \emptyset$, etc.

In an analogous way, one may show that the phase portrait of a
generic kCPD process,  such as, e.g., $kA\to \emptyset\,,\, kA\to
(k+1)A\,,\, (k+1)A\to kA$, contains a triangle of 
$k$--times--degenerate $q=0$ line, $p=0$ line [after the shift,
Eq.~(\ref{shift})], and the ``accidental'' $q=(m+up)/v$ diagonal
line; see Fig.~\ref{fig_kcpd}(b). Thus it may be described by a
model Hamiltonian of the form
\begin{equation}\label{kCPDham}
    H_R = (m+up-vq)\, p\,  q^k \, ,
\end{equation}
where $m$ is the control  parameter of the transition and  $u$ and
$v$ are positive constants.

\begin{figure}
\includegraphics[width=8cm]{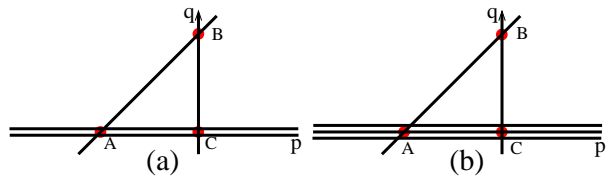}
\caption{Generic  phase portraits of (a) 2CPD models and (b) 3CPD
models in the active phase. }\label{fig_kcpd}
\end{figure}

To find scaling exponents near the transition one assigns bare
dimensions  $z=2$ and  $[p]+[q]=d$. There are no perturbative
corrections to the propagator (for $k\geq 2$), and thus one does
not expect these exponents to be changed in the
$\epsilon$ expansion. Since  both $u$ and $v$ vertices have to be
equally (ir)relevant on the mean-field level,  one has to choose
the bare dimensions as $[p]=[q]=d/2$. As a result, the bare
scaling dimensions of the vertices are $[m]=2-(k-1)d/2$, and
$[u]=[v]=2-kd/2$. According to this scaling analysis  the critical
dimension is expected to be $d_c=4/k$. The corresponding
mean-field exponents of the kCPD transitions at $d>d_c$ are
$\beta=1$ and $\nu_\bot=k/2$. Only the $k=2$ and $k=3$ processes are
expected to exhibit nontrivial behavior in the physically relevant
dimensions \cite{foot2,foot3}. We shall analyze these two cases
below.

The 2CPD transition is described by the action
(\ref{nonequilibrium}) with the reaction Hamiltonian (\ref{PCPDham})
(see, e.g., Ref.~[\onlinecite{HH04}]). Its critical dimension is
$d_c=2$. The one--loop renormalization is given by two--vertex loops,
which are logarithmically divergent in $d=2$. One arrives
\cite{JvWOT04} at the following set of RG equations:
\begin{eqnarray}\label{RGmuvPCPD}
  \partial_l m &=& (1+\epsilon/2 + \tilde S u)\,m\,, \label{m1} \\
  \partial_l u &=& (\epsilon + \tilde S u)\,u\,,\label{u1}\\
  \partial_l v &=& (\epsilon + 3 \tilde S u)\,v\, , \label{v1}
\end{eqnarray}
where $\epsilon=2-d$ and
\begin{eqnarray}\label{tildeSdp}
     \tilde S = \partial_l\frac{1}{2}\int_{\Lambda e^{-l}}^\Lambda \frac{\mathrm{d}^d k}{D k^2}\,
      = \ \frac{\Lambda^{d-2}}{4 \pi D}\,.
\end{eqnarray}
Notice that the sign of the perturbative corrections in
Eqs.~(\ref{m1})--(\ref{v1}) is {\em opposite} to that in the
corresponding DP RG equations (\ref{m})--(\ref{v}). As a result,
the weak--coupling fixed point appears to be absolutely unstable
for $d\leq d_c$ (for positive initial $u$). Its region of
stability for $d>d_c$ is finite and shrinking as $d\to d_c$ from
above \cite{footnew}. Solving Eq.~(\ref{u1}), one finds that the
coupling constant $u$ diverges once the RG reaches a certain
spatial scale $\xi^{-1}=\Lambda e^{-l}$, where, in $d<2$,
\begin{equation}\label{xi}
  \xi=\left({\epsilon D\over u_0}\right)^{1/(2-d)}
\end{equation}
(here $u_0$ is an initial value of $u$). In $d=2$ one finds that
$\xi=\Lambda^{-1}\exp(4\pi D/u_0)$. This indicates that some new
physics shows up at the scale $\xi$. In Appendix \ref{app2cpd} we
suggest that the system may develop anomalous averages, similar to
those in BCS theory. The corresponding ``coherence length''
appears to be exactly $\xi$.  (See also
Refs.~[\onlinecite{HH04,JvWOT04}] for further discussion.)

A similar situation is encountered for $k=3$. The critical
dimension is $d_c=4/3$. The RG analysis of Eq.~(\ref{kCPDham}) with
$k=3$ shows that the only vertex which acquires perturbative
corrections is $v$. The corresponding RG equations are
\begin{eqnarray}\label{RGmuvkCPD}
  \partial_l m &=& (2-d)\,m\,, \label{m2} \\
  \partial_l u &=& (2-3d/2)\,u\,,\label{u2}\\
  \partial_l v &=& (2-3d/2)\,v-3\tilde S mu\, . \label{v2}
\end{eqnarray}
Once again, the coupling constant $u$ grows indefinitely and the
$\epsilon$ expansion fails to predict critical exponents.


\subsection{Critical point and mean--field transitions}
\label{section_critpoint}

In the discussions above we have avoided considering an exact
location of the critical point in the parameter space of the
problem. Instead, we have concentrated on  the universal
properties of the transition itself. The only thing we had to
assume is that the three intersection points $A$, $B$, and $C$ in Figs.
\ref{fig_dp_gen} and \ref{fig_kcpd} can be made to collapse. In
other words, the parameter $m$ in Eqs.~(\ref{genericDP}) or
(\ref{kCPDham}) can be made to be zero. In terms of the bare
parameters of the model [e.g., Eq.~(\ref{DP})] this corresponds to
$\mu=k\lambda$. However, the actual transition may take place away
from the $\mu=k\lambda$ point. Indeed, it is the {\em
renormalized} parameter $m$ that vanishes at criticality, {\em
not} the bare one.

The particular case of bare $\lambda=0$, e.g.,
\begin{equation}\label{lambdazero}
  kA \stackrel{\mu}{\rightarrow} (k+1) A\,, \quad \,\,
    (k+1) A \stackrel{\sigma}{\rightarrow} \emptyset\
\end{equation}
has attracted a lot of attention \cite{CardyTauber,NPRG,Odor04a}.
The naive mean--field expectation is that such a system is always
in its active phase, unless $\mu=0$. Yet it is clear that any
finite--size system is bound to end up in its empty absorbing state
after long enough time. [The latter may be estimated as an
exponentiated classical action accumulated along the trajectory
leading from the active mean-field state (point $B$ in Figs.
\ref{fig_dp} and \ref{fig_pcpd0}) to the empty absorbing state
(point $A$) \cite{ElgartKamenev04}.] Therefore, if the diffusion
time between the sites is longer than the respective extinction
time, the system evolves towards extinction even at $\mu >0$. As a
result, there is a nontrivial phase separation boundary in the
parameter space $(D,\mu)$ of the $\lambda=0$ system \cite{NPRG},
Fig.~\ref{fig_boundary}. One may interpret the appearance of this
nontrivial line as the generation of the effective annihilation
rate $\lambda_{eff}(D,\mu)>0$. For example, two successive
processes $A\rightarrow 2A$ and $2A\rightarrow\emptyset$ result in
an $A\rightarrow\emptyset$ effective rate. Consequently the phase
transition along the nontrivial critical line is represented by
the evolution of the phase portraits depicted in
Figs.~\ref{fig_dp_gen} and \ref{fig_kcpd} and belongs to the kCPD
universality classes.

\begin{figure}
\includegraphics[width=8cm]{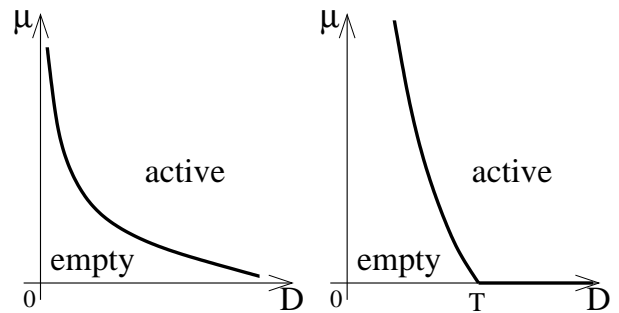}
\caption{Schematic phase boundary in the parameter space $(D,\mu)$
of the $\lambda=0$ reaction set (\ref{lambdazero}). (a) $d\leq
d_c'$, the transition along the entire line is of kCPD class. (b)
$d>d_c'$, to the left of the tricritical point $T$ the transition is
of kCPD class, while to the right it is $\mu=0$ mean-field
transition.}\label{fig_boundary}
\end{figure}

The crucial observation made by Cardy and T\"auber
\cite{CardyTauber} is that at and below a certain lower critical
dimension $d_c'$ the separation boundary extends to an arbitrarily
large diffusion constant $D$, Fig.~\ref{fig_boundary}(a). This
fact may be associated with a divergent perturbative correction to
the coupling constant $m$. In the two most interesting cases DP
and 2CPD such a correction is given by the two-vertex loop,
Eq.~(\ref{tildeSdp}). As a result, the corresponding lower
critical dimension is $d_c'=2$. Above the lower critical
dimension, $d>d_c'$, there is  a tricritical point $T=(D_c,0)$,
such that for $D>D_c$ the transition occur only at $\mu=0$
\cite{NPRG}, Fig.~\ref{fig_boundary}(b).

\begin{figure}
\includegraphics[width=8cm]{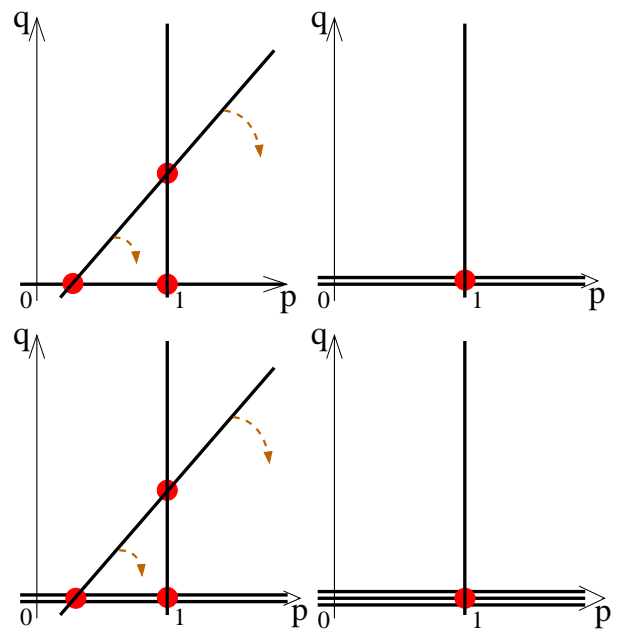}
\caption{(Color online) Phase portrait evolution upon $\mu=0$
mean--field transitions: (a) $k=1$ and $\mu>0$, decreasing  the
branching rate $\mu$  is schematically shown by the  dashed
arrows. (b) $k=1$ and $\mu =0$, the ``accidental'' line becomes
horizontal and the local topology of the phase portrait is that of
the binary annihilation reaction; cf. Fig.~\ref{fig_ann}. (c)
$k=2$ and $\mu>0$, (d) $k=2$ and $\mu=0$, the topology is that of
the $3$-plet annihilation reaction, $3A\rightarrow\emptyset$.}
\label{fig_mf}
\end{figure}

The evolution of the phase portrait corresponding to such a
$\mu=0$ transition is depicted in Fig.~\ref{fig_mf} (cf.
Figs. \ref{fig_dp_gen} or \ref{fig_kcpd}). At $\mu=0$ the only
remaining reaction is $n\geq 2$ particle annihilation, $(k+1)A
\rightarrow \emptyset$, with the critical dimensionality
$d_c=2/(k-1)\leq 2$. Thus, the critical dimensionality of the
$\mu=0$ transitions is smaller than $d_c'$ -- the minimum
dimensionality above which the $\mu=0$ transitions may take place.
As a result, all such transitions occur {\em above} their
respective critical dimensionality and therefore exhibit
mean-field behavior. The generic scheme for a mean-field $\mu =0$
transition may be written as \cite{Odor03}
\begin{equation}
    k A \stackrel{\mu}{\rightarrow} (k+l) A\,,\label{mu_zero}\quad\,\,
    n A \stackrel{\sigma}{\rightarrow} (n-m) A\, ,
\end{equation}
where $k<n$. The corresponding rate equation is
\begin{equation}\label{mu_zero_rate}
  \partial_t q= \mu\, {l\over k!}\, q^k - \sigma\, {m\over n!}\,
  q^n\, .
\end{equation}
The stationary density is proportional $\langle n\rangle \propto
(\mu/\sigma)^{1/(n-k)}$, leading to the mean-field critical exponent
$\beta=1/(n-k)$. The other mean-field exponents can be also deduced
in a straightforward way \cite{Odor03}.

There are two important exceptions where the $\mu=0$ transitions
in the reaction set Eq.~(\ref{mu_zero}) may exhibit non-mean-field
behavior. These are the cases where  generation of the effective
$\lambda_{eff}(D,\mu)$ is prohibited by certain symmetries. This
may lead to a scenario with $d_c' < d_c$, thus leaving a window
$d_c' < d \leq d_c$ for non--mean--field $\mu=0$ transitions.
There are two such symmetries: (i) parity conservation (both $l$
and $m$ are even), discussed in Sec.~\ref{section_PC}, and (ii)
reversal symmetry ($k+l=n$ and $n-m=k$), discussed in Sec.~\ref{sec_rev}.

\section{Parity Conserving models}\label{section_PC}

\subsection{PC model}
\label{secPC}

As was mentioned above the parity conservation may lead to a
new nontrivial universality class.  Consider the simplest
possible PC reaction set
\begin{equation}
   A \stackrel{\mu}{\rightarrow} 3 A \,, \quad\,\,\,
    2 A \stackrel{\sigma}{\rightarrow} \emptyset\,.\label{PC}
\end{equation}
The corresponding reaction Hamiltonian takes the form
\begin{equation}\label{hamiltonianPC}
    {H}_{R} = \left(u p-v\,q\right)(p^2-1)\,{q}\, ,
\end{equation}
where the bare values of the constants are given by $u=\mu$ and
$v=\sigma/2$.  The corresponding action  is invariant under the
following transformations \cite{CardyTauber}:
\begin{equation}
    p \rightarrow -p \,,\quad\,\,\,
    q \rightarrow -q \,,
 \label{sym_pc}
\end{equation}
which may be traced back to the conservation of parity. As a
result, the phase portrait, Fig.~\ref{fig_pc}, possesses the
reflection point at the origin. This symmetry is preserved upon RG
transformations. Therefore the ``accidental'' zero-energy line
[$q=u p/v$,  according to Eq.~(\ref{hamiltonianPC})] is bound to
be an odd function and thus pass through the origin. Its shape,
however, may change in the process of renormalization.
Consequently  the phase transition  cannot be described by the
coalescence of  three points $A$, $B$, and $C$ and its nature is different
from the DP class.

\begin{figure}
\includegraphics[width=8cm]{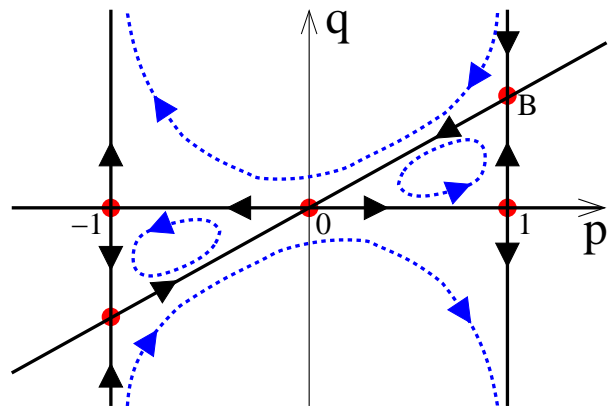}
\caption{(Color online) Phase portrait of the parity conserving
  model, Eq.~(\ref{PC}). Notice the reflection symmetry around the origin.}\label{fig_pc}
\end{figure}

According to the mean-field equation $\partial_t
q=2\mu\,q-\sigma\, q^2$, the model is always in the active phase
with the number of particles  $\langle n\rangle =
2\mu/\sigma=u/v$. The  only way to drive the mean-field dynamics
towards extinction is to send $u=\mu\to 0$. In other words,
the critical point is $u_c=0$. One may discuss, though, the
scaling of particle density with $u-u_c=u$ and this way define the
``magnetization'' exponent $\beta$:
\begin{equation}\label{beta}
  \langle n\rangle\sim u^\beta \, ,
\end{equation}
where the mean-field value of the exponent is $\beta=1$.

Turning to the fluctuations,  one notices that it is not possible
to perform the shift of momentum, Eq.~(\ref{shift}), and focus on
the immediate vicinity of the $(1,0)$ point on the phase plane.
Because of the symmetry (\ref{sym_pc}) one has to keep the entire
interval $p \in[-1,1]$ under consideration; see Fig.~\ref{fig_pc}.
Therefore one must choose the scaling dimension $[p]=0$. Since the
bare scaling requires $[p]+[q]=d$, the naive scaling dimension of
$q$ is $[q]=d$. As a result, one finds (since $z=2$) $[u]=2$ and
$[v]=2-d$ and the critical dimension is $d_c=2$. One may  worry
that since $[p]=0$, it is not possible to restrict the
consideration to the low--order polynomial in $p$. Instead, one has
to keep all the powers of $p$, resorting to the functional RG. We
perform such a procedure in Appendix \ref{appFRG} and show that it
actually justifies the use of the truncated reaction Hamiltonian
(\ref{hamiltonianPC}).

The one--loop RG calculation, utilizing the two-vertex loop (which
is logarithmically divergent in $d=2$), yields  the following RG
equations \cite{CardyTauber,THL05}:
\begin{eqnarray}
    \partial_l u &=& (2  - 3\tilde S\,v)\,u \,,\label{u3}\\
    \partial_l v &=& (\epsilon - \tilde S\,v)\,v\,,\label{v3}
\end{eqnarray}
where $\epsilon=2-d$ and $\tilde S$ is given by
Eq.~(\ref{tildeSdp}). For $\epsilon>0$ there is a nontrivial stable
fixed point $v^*=\epsilon/\tilde S$. In the vicinity of this fixed
point the  relevant parameter $u$ scales as $\partial_l u = (2 -
3\epsilon)\,u$ and thus its new scaling dimension is
\begin{equation}\label{uscaling}
  [u]=2-3\epsilon=3d-4\, .
\end{equation}
This leads to the  non-mean-field exponent $\beta$ given by
\begin{equation}\label{beta1}
  \beta=d/[u]\approx 1+\epsilon\, .
\end{equation}
The fact that $\beta>\beta^{MF}=1$ means that the actual density
in $d<2$ is less than the mean-field prediction. The fluctuations
drive the system closer to extinction. Cardy and T\"auber
\cite{CardyTauber} suggested that for $d<d\,'_c\approx 4/3$  there
is an extinct phase  at finite $\mu$ and the transition to the
active phase takes place at some $\mu_c>0$ and estimated the
corresponding critical exponents. Recently, the PC model was
reexamined employing the so--called nonperturbative
renormalization group method by Canet {\em et al.} 
\cite{CCDDM05,Canet06}. They confirmed the existence of the
$\mu>0$ transition for the PC model below $d\,'_c=4/3$ and
estimated the exponents with fair accuracy for $d=1$.

\subsection{Generalized PC models}

One may  invent other models conserving parity. For example, Cardy
and T\"auber \cite{CardyTauber} considered the class of parity
conserving models $2A\to \emptyset$, and $A\to (m+1)A$ with even
$m$. The corresponding reaction  Hamiltonian is
$H_R=[u p\,h_m(p)-v\,q]\,(p^2-1)\,q$, where $h_m(p)=
(p^m-1)/(p^2-1)$ is an even polynomial. Its phase portrait is
topologically identical to Fig.~\ref{fig_pc}. Thus one expects
this reaction set to belong to the same universality class as the PC
model. This was indeed the conclusion of
Ref.~[\onlinecite{CardyTauber}].

To find a different topology of the phase portrait and therefore a
new universality class one needs to impose an additional symmetry.
Following Sec.~\ref{seckcpd}, we shall consider parity
conserving reactions with minimal number $k>1$ needed to initiate
all reactions. For example, consider a parity conserving set of
reactions with even $k$:
\begin{equation}
   kA \stackrel{\mu}{\rightarrow} (k+2) A \,,\quad\,\,\,
    (k+1) A \stackrel{\sigma}{\rightarrow} A\,,\label{kPCeven}
\end{equation}
The reaction Hamiltonian is
\begin{equation}\label{hamiltoniankPC}
    {H}_{R} = \left[u p^{k-1}-v\,h_{k}(p)\,q\right]p(p^2-1)\,q^k\, ,
\end{equation}
where $h_{k}(p)$ is an even polynomial of  the degree $k-2$,
$u=\mu/k!$, and $v=\sigma/(k+1)!$. The corresponding phase portrait is
depicted in Fig.~\ref{fig_pcgen}(a). For an odd $k$ a
representative set of reactions is
\begin{equation}
   kA \stackrel{\mu}{\rightarrow} (k+2) A \,,\quad\,\,\,
    (k+1) A \stackrel{\sigma}{\rightarrow} \emptyset\,.\label{kPC}
\end{equation}
with the reaction Hamiltonian
\begin{equation}\label{hamiltoniankPCodd}
    {H}_{R} = \left[u p^k-v\,h_{k+1}(p)\,q\right](p^2-1)\,q^k\, ,
\end{equation}
where $h_{k+1}(p)$ is an even polynomial of degree $k-1$ . The
corresponding phase portrait is depicted in
Fig.~\ref{fig_pcgen}(b). These phase portraits are topologically
stable upon RG transformations and thus represent a set of
distinct universality classes. We call them $k$-particle parity
conserving (kPC) classes.

\begin{figure}
\includegraphics[width=8cm]{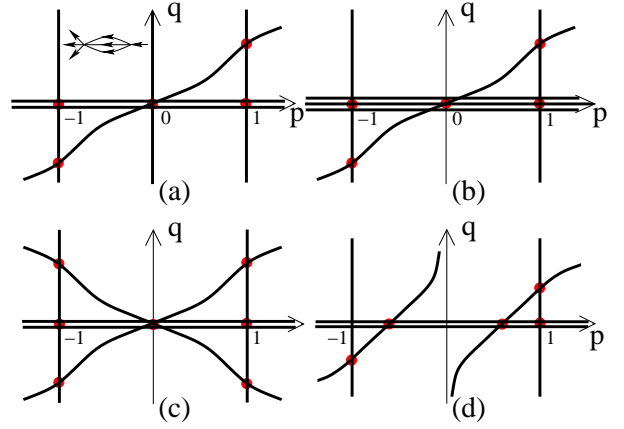}
\caption{Phase diagrams of kPC models. (a) 2PC model, the inset
shows the diagram leading to the logarithmic corrections in $d=1$.
(b) 3PC model. (c) 2PC model with all reactions including only an even
number of particles. This topology is not stable and evolves towards
(a). (d) 2PC with $2A\to \emptyset$, the corresponding topology is
essentially equivalent to the 2CPD model, Fig.~\ref{fig_pcpd0}.
}\label{fig_pcgen}
\end{figure}

Assigning the scaling dimensions as in the PC model, $[p]=0$,
$[q]=d$, and $z=2$, one finds $[u]=2-(k-1)d$ and $[v]=2-kd$. At the
critical dimension $v$ turns out to be relevant and thus $d_c=2/k$.
Therefore at any physical dimension the kPC behavior is described
by the mean--field. The only exception is the 2PC model,
Fig.~\ref{fig_pcgen}(a), which acquires logarithmic corrections to
the mean-field scaling at $d=1$. The renormalization is due to the
two-loop diagrams built with the help of the $p^3q^3$ vertex; see the
inset in Fig.~\ref{fig_pcgen}(a).

Other attempts to generalize the PC universality class appear to
be unstable against RG transformations. For example, consider a
parity conserving set which contains only an {\em even} number of
reagents: $2A\to 4A$ and $4A\to \emptyset$. The corresponding
reaction Hamiltonian is given by
$H_R=[up^2-v(1+p^2)q^2](p^2-1)q^2$. Its phase portrait is depicted
in Fig.~\ref{fig_pcgen}(c). In addition to the PC symmetry,
Eq.~(\ref{sym_pc}), the Hamiltonian and the phase portrait possess
\begin{equation}\label{q-q}
  q\to -q\,
\end{equation}
symmetry. However, this is {\em not} the symmetry of the full
action, Eq.~(\ref{nonequilibrium}). Therefore this symmetry is not
stable against the RG transformations. Indeed, e.g., using three
vertexes $p^2q^2$, one may generate a $p^3q^3$ vertex which violates
the symmetry (\ref{q-q}). As a result the system belongs to the
2PC class and its phase portrait drifts towards
Fig.~\ref{fig_pcgen}(a).

One may add a competing annihilation  reaction to the $kA\to
(k+2)A$ process of kPC for $k\geq 2$, such as $kA\to (k-2)A$. For
example,
\begin{equation}
   2A \stackrel{\mu}{\rightarrow} 4 A \,,\quad\,\,\,
     3A \stackrel{\sigma}{\rightarrow} A\,,\label{kPCevenPCPD}
     \quad\,\,\,
   2A \stackrel{\lambda}{\rightarrow} 0 \, .
\end{equation}
Because of the competition, one expects the absorbing state
transition to happen at $m=2\mu-\lambda=0$. The Hamiltonian is
$H_R=(m-u+up^2-vpq)(p^2-1)q^2$; it obeys the PC symmetry
(\ref{sym_pc}). The phase portrait is plotted in
Fig.~\ref{fig_pcgen}(d). One notices that in the vicinity of the
transition the {\it local} topology is indistinguishable from
Figs.~\ref{fig_pcpd0} and \ref{fig_kcpd}(a). Therefore the
transition belongs to the same universality class as 2CPD (see
Sec.~\ref{seckcpd}). This fact was already noticed in numerical
simulations \cite{Park01,Odor02a}, but, to the best of our
knowledge, remained unexplained. The identity of the phase
portrait's topologies  in the vicinity of transition immediately
explains the universality. One can show that the other kPC
processes with competition in the $k$-particle channel belong
to the same universality classes as the corresponding kCPD models. The
example of the model, Eq.~(\ref{kPCevenPCPD}), shows that parity
conservation is not a crucial feature to discriminate the PC
class (see also Refs.~[\onlinecite{Hinrichsen00,HCDM05}]). It is
rather the topology of the phase portrait in the vicinity of the
transition that discriminates various universality classes.

\section{Reversible reactions}\label{sec_rev}

All the phase transitions, discussed above, are associated with a
deformation and rearrangement of a characteristic {\em triangular}
structure (possibly with one degenerate side) on the phase plane.
There is one more possibility  for a stable transition
topology which is a  {\em rectangular} structure. All models
exhibiting rectangular topology consist of a {\em single}
reaction which is allowed to go in both directions with different
rates.

Consider, for example, the reversible reaction [same as DP,
Eq.~(\ref{DP}), but without $A\to 0$]
\begin{equation}
    A \stackrel{\mu}{\rightarrow} 2 A\,, \label{revers}\quad\,\,
    2 A \stackrel{\sigma}{\rightarrow} A\,.
\end{equation}
The corresponding reaction Hamiltonian is
\begin{eqnarray}\label{hamiltonianrev}
    {H}_{R} &=&  \mu (p^2-p)\,{q}+ {\sigma\over 2} (p-p^2)\,{q}^2\nonumber \\
    &=&   (p^2-p)(u\,q-v\,q^2)\, ,
\end{eqnarray}
where $u=\mu$ and $v=\sigma/2$. The phase portrait is depicted in
Fig.~\ref{fig_rev}(a) and has characteristic rectangular shape
comprised by the generic lines $p=1$ and $q=0$ along with the {\em
two} ``accidental'' ones $p=0$ and $q=u/v$. The mean--field
predicts the average density to be $\langle n\rangle =u/v$. The
control parameter is $u=\mu$ with the critical value $u_c=0$.
Consequently the mean-field ``magnetization'' exponent is
$\beta=1$, in agreement with the general scheme \cite{Odor03};
cf.  Eqs.~(\ref{mu_zero}) and (\ref{mu_zero_rate}). One may ask
if the mean-field behavior  can be modified by the fluctuations.

\begin{figure}
\includegraphics[width=8cm]{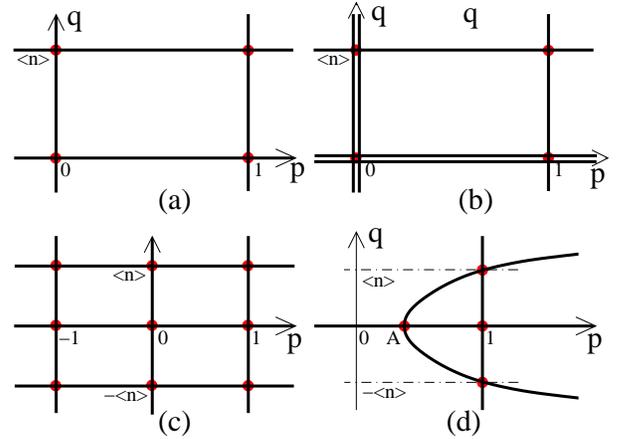}
\caption{(a)--(c) Phase portraits of reversible reactions. (a)
$A\leftrightarrow 2A$ and (b) $2A\leftrightarrow 3A$. (c) Parity
conserving $A\leftrightarrow 3A$. (d) A  topology with $q\to -q$
symmetry unstable against RG transformations; see 
Sec.~\ref{section_conclusion}. }\label{fig_rev}
\end{figure}

To answer this question, one notices that the phase portrait
topology is stable against renormalization. I.e., no terms
violating the rectangular structure are generated. It may be
checked either by considering possible diagrams or realizing that
the action possesses the symmetry
\begin{equation}\label{sym_rev}
  p \rightarrow {v\over u}\, q\, , \quad\,\,\,  q \rightarrow {u\over v}\,p\, , \quad\,\,\,
t \rightarrow -t\, .
\end{equation}
To keep the entire interval $p\in[0,1]$ unchanged upon
rescaling \cite{JMS}, one has to choose the scaling dimensions as
$[p]=0$ and $[q]=d$, then $[u]=2$ and $[v]=2-d$. From here one
concludes that the critical dimension is $d_c=2$. The RG equations
are
\begin{eqnarray}
    \partial_l u &=& (2  - \tilde S\,v)\,u \,,\label{u4}\\
    \partial_l v &=& (\epsilon - \tilde S\,v)\,v\,,\label{v4}
\end{eqnarray}
where $\epsilon=2-d$ and $\tilde S$ is given by
Eq.~(\ref{tildeSdp}). For $\epsilon>0$ there is a  stable fixed
point at $v^*=\epsilon/\tilde S$. In its  vicinity $u$ scales as
$\partial_l u = (2 - \epsilon)\,u$; thus its new scaling dimension
is $[u]=2-\epsilon=d$ (see also Appendix \ref{appFRG}). As a
result, the exponent  is given by $\beta=d/[u] = 1+O(\epsilon^2)$.
At least in this order, it is not affected by the fluctuations. In
fact,  $\beta=1$ is proven  to be exact in $d=1$,
Ref.~[\onlinecite{bAH00}].  Using detailed balance,
Ref.~[\onlinecite{JMS}] argues that $\beta =1$ is exact in any
dimension.

A generic reversible reaction $kA\leftrightarrow mA$, with $m>k$, is
described by  $H_R=(p^m-p^k)(uq^k-vq^m)$. Its zero-energy lines are
$q=0$ and $p=0$, both $k$ times degenerate, along with
nondegenerate $p=1$ and $q=\langle n\rangle=(u/v)^{1/(m-k)}$. For
the parity conserving case ($m-k$ even) also $p=-1$ and $q=-\langle
n\rangle$, Fig.~\ref{fig_rev}(c). By a proper rescaling of $p$ and
$q$ the Hamiltonian may be brought to the symmetric separable form
$H_R=-f(p)f(q)$. The corresponding action is  symmetric against
$p\leftrightarrow q$ and $t\to -t$. Therefore the rectangular
structure is stable in the course of renormalization. The
topological structure is fully determined by the index $k$ and the
parity.

Therefore, one may identify two more families of universality
classes. One is parity nonconserving, represented by the
reversible reaction
\begin{equation}
    kA \leftrightarrow (k+1) A\,  \label{krevers}
\end{equation}
(a higher number of offsprings, $k+1+2n$, does not change the
universality class) with the reaction Hamiltonian
\begin{equation}\label{hamiltoniankrev}
    {H}_{R} =(p-1)(u-v\,q)p^k\,q^k\, .
\end{equation}
We denote it as $k$-particle reversible (kR). The action possesses
the symmetry (\ref{sym_rev}), rendering stability of the
rectangular topology.  The upper  critical dimension is $d_c=2/k$.
An example is $2A\leftrightarrow 3A$ [see Fig.~\ref{fig_rev}(b)],
with $d_c=1$.

The parity conserving reversible reactions
\begin{equation}
    kA \leftrightarrow (k+2) A\,  \label{kreversPC}
\end{equation}
(a higher number of offsprings, $k+2n$, does not alter the
universality class) with the reaction Hamiltonian
\begin{equation}\label{hamiltoniankrevPC}
    {H}_{R} =(p^2-1)(u-v\,q^2)p^k\,q^k\, .
\end{equation}
We denote it as kRPC. The corresponding action is symmetric
against two symmetries, Eqs.~(\ref{sym_pc}) and (\ref{sym_rev}).
They impose stability of the rectangular topology symmetric with
respect to reflection around the origin. The critical dimension of the
kRPC family is $d_c=2/(k+1)$. An example is $A\leftrightarrow 3A$
[see Fig.~\ref{fig_rev}(c)], with $d_c=1$. All other reactions of
this type have  $d_c<1$ and thus are fully described by the
mean-field treatment (see, however, Ref.~[\onlinecite{Odor03}]).

\section{Conclusions}
\label{section_conclusion}

We have argued that the universality classes of phase transitions
in reaction-diffusion models may be classified according to the
topological structures of the corresponding phase spaces. This
structure   is fully encoded in the web of the zero-energy
trajectories. The simplest and most stable structure is given by
three mutually intersecting  lines. By changing a single control
parameter the three intersection points  may be made to coalesce;
see Fig.~\ref{fig_dp_gen}. At such a value of the control
parameter the system undergoes a phase transition into the
absorbing phase. The corresponding universality class is known as
directed percolation.

There are only a limited number of ways to organize a {\em
continuous} phase transition, governed by a {\em single} control
parameter, which utilizes topology different from the DP. We have
identified {\em four} families of such transitions, which are
stabilized by an additional symmetry or symmetries.

(i) A generic  reaction set constrained by the requirement that
{\em all} reactions need at least  $k$ incoming particles. The
corresponding phase portrait is bound to have the $q=0$ line to be
$k$ times degenerate, Fig.~\ref{fig_kcpd}. This property is robust
against RG transformations. Indeed, no vertices  with fewer than
$k$ external $q$ ``legs'' can be generated. The triangular
topology with one  $k$-degenerate line, Fig.~\ref{fig_kcpd},
defines a family of universality classes. We call them kCPD's
($k=1$ is DP). The scaling considerations suggest that their upper
critical dimension is $d_c=4/k$ \cite{footnew}.

(ii) A set of reactions which conserve parity. In this case the
Hamiltonian and the action are invariant under the transformation
(\ref{sym_pc}). It dictates the reflection symmetry of the
corresponding phase portraits, Figs.~\ref{fig_pc} and
\ref{fig_pcgen}. The symmetry is preserved upon renormalization.
In addition to parity conservation one may require a minimal number
$k$ of incoming particles for every reaction. This generates the phase
portraits depicted in Fig.~\ref{fig_pcgen}. There is one
universality class for every $k$, termed kPC,
Figs.~\ref{fig_pcgen}(a) and \ref{fig_pcgen}(b) ($k=1$ is PC). Their upper  critical
dimension is $d_c=2/k$. To realize a kPC transition, the reaction
starting from the minimal number $k$ of particles must go only up,
e.g., $kA\to (k+2)A$. By adding down--going reaction, e.g., $kA\to
(k-2)A$ (for $k\geq 2$) -- the model is transformed into the kCPD
class (despite of the parity conservation).

(iii) A single reaction which is allowed to go both directions,
with different rates: $kA\leftrightarrow (k+1+2n)A$. The
corresponding reaction Hamiltonian and the action are symmetric
under the exchange transformation, Eq.~(\ref{sym_rev}). The phase
portrait has the stable rectangular structure with $k$-degenerate
$p=0$ and $q=0$ lines, Figs.~\ref{fig_rev}(a) and \ref{fig_rev}(b). Upon decreasing
the creation rate  the rectangle collapses onto the interval
$p=[0,1]$. We call such transitions kR. Their critical dimension
is $d_c=2/k$ (same as kPC, but exponents are different in
$d<d_c$).

(iv) A single reaction which is allowed to go both directions and
conserves parity: $kA\leftrightarrow (k+2n)A$. The corresponding
reaction Hamiltonian and the action are symmetric under the {\em
two} symmetry transformations, Eqs.~(\ref{sym_pc}) and
(\ref{sym_rev}). The corresponding phase portrait has the stable
``checkered'' structure, Fig.~\ref{fig_rev}(c), which  collapses
when sending the creation rate to zero. We denote such transitions as
kRPC. Their critical dimension is $d_c=2/(k+1)$.

Altogether we identify five nontrivial universality classes with
$d_c>1$: DP, PCPD, 3CPD, PC, and 1R. In addition there are four
marginal classes with $d_c=1$: 4CPD, 2PC, 2R, and 1RPC. In
sufficiently high  dimensions there are other universality
classes not belonging to the four families described above.
However, they are bound to be of the {\em mean--field} type and do
not have nontrivial representatives.

We have not found any other stable, topologically distinct
structures in the phase space. Consider, for example, the topology
depicted in Fig.~\ref{fig_rev}(d). The phase portrait is symmetric
under the transformation $q\to -q$. A corresponding reaction set
consists of reactions {\em all}  starting from {\em odd} number of
particles: e.g. $A\to \emptyset$, $A\to 2A$, and $3A\to
\emptyset$. The corresponding Hamiltonian [after the shift
(\ref{shift}) and neglecting irrelevant terms] is
$H_R=(m+up-vq^2)\,p\,q$.
By changing the parameter $m$, e.g., by
increasing the rate of annihilation $A\to \emptyset$ -- one may
bring the system to the phase transition into the extinction
phase. Naively, such a transition is associated with the vertex of
the parabola crossing the point $(1,0)$; see
Fig.~\ref{fig_rev}(d). Such a topology is different from all those
considered above and could represent a new universality class.

However, the $q\to -q$ symmetric phase portrait,
Fig.~\ref{fig_rev}(d), is {\em unstable} upon renormalization.
Indeed, combining two vertices  $p\,q^3$ and $p^2q$ in the loop,
one generates the $p\,q^2$ term, which violates the symmetry. This
term represents the induced reaction $2A\to A$, originating from
two reactions $A\to 2A$ followed by $3A\to\emptyset$. In other
words, $q\to -q$ is {\em not} a symmetry of the action and
therefore it is not preserved by the RG.  As a result, the
initially symmetric zero-energy line $p=(vq^2-m)/u$ is shifted and
deformed upon renormalization. The topology  drifts either towards
the first--order scenario, Fig.~\ref{fig_fot_II}, or towards the
DP, Fig.~\ref{fig_dp_gen}.  To keep the vertex of the parabola
right at  the $q=0$ line, one has to fine-tune at least one
additional control parameter, besides $m$. This is the case of a
tricritical transition point
\cite{Janssen87,OK87,JMS04,Lub06,Grassberger06}.

In the present work we have restricted ourselves to the
one--component reactions. It is desirable  to extend the strategy
to reaction--diffusion models with several reagents. Each new
reagent brings an additional reaction coordinate and corresponding
momentum. E.g., a two--component model requires four dimensional phase
space with $d=3$ surfaces of constant energy. The corresponding
classical dynamics may be not integrable, making the phase--space
topology rather complicated. The situation may be simplified by
the presence of  conservation laws. For example, reaction $A+B\to
2B$ conserves the number of particles. This leads to a classical
dynamics with an additional integral of motion, besides energy. It
is most clearly seen after a canonical transformation
\cite{Jordan} $p=e^P$, $q=Q\,e^{-P}$, which leads to the integral
of motion $Q_A+Q_B=\mbox{const}$. Even with such simplifications
understanding the full phase--space dynamics of multicomponent
systems remains a challenge.

\subsection*{Acknowledgments}

The authors thank B.~Delamotte, I.~Dornic, H.~Hinrichsen,
S.~L{\"u}beck, M.~Mu\~noz, G.~\'Odor and U.~T{\"a}uber for useful
communications and discussions after online publication. This work was supported by
NSF Grant No.~DMR-0405212 and by the A.~P.~Sloan Foundation.

\appendix

\section{2CPD model}
\label{app2cpd}

Here we consider the 2CPD model (same as PCPD) which was shown to
exhibit runaway RG flow for $d\leq 2$; see
Sec.~\ref{seckcpd}. We shall argue that this behavior of the RG
indicates a rearrangement of the ground state, such that the vacuum
supports the anomalous averages of the type $\langle q^2 \rangle$,
similar to the BCS theory.

In a vicinity of the phase transition the action of the model is
given by Eq.~(\ref{nonequilibrium}) with the reaction  Hamiltonian
(\ref{PCPDham}). To ensure convergence of the functional integral,
it is convenient to perform rotation of the integration contour in
the complex $p$ plane: $p\to ip$. This way one arrives at an action
of the form
\begin{equation}\label{act2cpd}
    S = \!\int \!\!dt\, d^d x\, \Big[ip\,
    \left(\partial_t q - D\nabla^2q - m \, q^2   + v\,q^3\right) + up^2q^2\Big].
\end{equation}
We note, in parentheses, that this  is the Martin-Siggia-Rose
action of the following Langevin  process with the multiplicative
noise $\eta$:
\begin{equation}\label{langevin}
\partial_t q = D\nabla^2q + m \, q^2   - v\,q^3 + q\,\eta(x,t)\, ,
\end{equation}
where $\langle \eta(x,t)\eta(x',t')\rangle
=2u\delta(x-x')\delta(t-t')$.

We shall assume now that the vacuum of the theory supports the
anomalous average value
\begin{equation}\label{Delta}
  \langle q^2(x,t)\rangle =\Delta\, ,
\end{equation}
which is to be determined from the self--consistency condition.
Neglecting the nonlinear fluctuation terms, the action acquires
the form
\begin{equation}\label{actGauss}
    S = \!\int \!\!dt\, d^d x\, \Big[
    ip\left(\partial_t  - D\nabla^2   + v\Delta\right)q + u\Delta p^2 -im\Delta p \Big].
\end{equation}
With this Gaussian action one can evaluate $\langle q^2(x,t)\rangle$
and impose the  condition (\ref{Delta}). This leads to the
self--consistency  equation
\begin{equation}\label{selfconsist}
  \Delta=u\Delta \int\!\! \frac{\mathrm{d}^d k}{D k^2+v\Delta}\,
  +\,  \frac{m^2}{v^2}\, .
\end{equation}
Without the first term on the RHS, $\Delta=(m/v)^2$, this is
simply the mean--field prediction for $\langle q^2(x,t)\rangle$. In
$d\leq 2$ this equation has a nontrivial solution even at $m=0$.
In particular at $d=2$ one finds
\begin{equation}\label{Delta2}
  \Delta^{(d=2)} = {\Lambda^2 D\over v}\,\, e^{\, -4\pi D/u}\, ,
\end{equation}
where $\Lambda\sim 1/a$ is the momentum cutoff. In $d<2$ one obtains
\begin{equation}\label{Delta<2}
  \Delta^{(d<2)} \sim { D\over v}\,\left({u\over (2-d) D}\right)^{2/(2-d)}\, .
\end{equation}
As a result, one finds that the ``order parameter'' $\Delta$ is
associated with the ``coherence length'' $\xi$, given by
\begin{equation}\label{vDelta}
  v\Delta=D\,\xi^{-2}\, .
\end{equation}
Notice that $\xi$ is exactly the characteristic spatial scale for
the breakdown of the RG treatment of Sec.~\ref{seckcpd}; see
Eq.~(\ref{xi}). This consideration  suggests that the divergence
of the RG flow is associated with the development of the anomalous
average, Eq.~(\ref{Delta}).

Let us mention a two--field coupled Langevin description that has
been proposed for the related pair contact process without
diffusion \cite{MGDL96} and for 2CPD model as well \cite{DCM05a}.
In both cases a pair field and a singlet field were
introduced in order to show a phase transition and estimate the
critical exponents (in fair agreement with those reported for
microscopic models). More work is needed to appreciate the connections
(if any) with our anomalous average microscopic approach.

\section{Functional Renormalization Group}
\label{appFRG}

In some problems one cannot focus on the immediate vicinity of
the $(1,0)$ point in the phase plane $(p,q)$. Instead, one has to
keep under consideration the entire interval $p\in[0,1]$, or even
$p\in [-1,1]$. This happens, e.g., in  parity conserving models,
because of the $p\to -p$, $q\to -q$ symmetry. To keep the
$p$ interval intact upon renormalization, one must choose the
scaling dimension $[p]=0$. Since the bare scaling requires
$[p]+[q]=d$, one is left with the naive scaling dimension $[q]=d$.
With such scaling dimensions one may restrict the reaction
Hamiltonian to the lowest powers in $q$ (typically the first and
second), but one must keep {\em all}  powers of $p$. As a result,
one has to employ the functional RG treatment (cf.
Ref.~[\onlinecite{JvWOT04}]).

The generic reaction Hamiltonian for the absorbing-state models,
mentioned above, is
\begin{equation}\label{effectPC}
    H_{R} = f(p)\,q - g(p)\,q^2 \, ,
\end{equation}
where $f(p)=\sum_n f_n p^n$ and $g(p)=\sum_n g_n p^n$ are
polynomials of $p$. From the normalization condition (\ref{norm2})
it follows that
\begin{equation}\label{pm1}
  f( 1)=g( 1)=0\, .
\end{equation}
Specific models may possess additional symmetries which dictate
further restrictions on the polynomials $f(p)$ and $g(p)$; e.g.,
for PC models $f(p)$ is odd, while $g(p)$ is even, due to
parity conservation: $p\to -p$, $q\to -q$. We shall keep the
presentation general, imposing these additional symmetries at a
latter stage. Because of the assigned bare scaling dimensions
$[p]=0$, $[q]=d$, and $z=2$, one finds $[f_n]=2$ and $[g_n]=2-d$.
Thus the formal critical dimension is $d_c=2$.

The one loop renormalization is given by the two-vertex loops and
leads to the following set of the RG equations:
\begin{eqnarray}
    \partial_l f_n &=& 2 f_n - \frac{1}{2}\tilde S \sum_{m,k}m(m-1)\, \delta_{n,m+k-2}f_m g_k \,,\label{f_n}\\
    \partial_l g_n &=& \epsilon g_n -\frac{1}{2}\tilde S \sum_{m,k}m(m-1)\,\delta_{n,m+k-2}g_m g_k\,,\label{g_n}
\end{eqnarray}
where $\epsilon = 2-d$ and $\tilde S$ is given by
Eq.~(\ref{tildeSdp}). The factors $\frac{1}{2}m(m-1)$  describe the combinatorial
number of pairs which may form  the loop. The $\delta$ symbols
enforce the proper number of the external (slow) ``legs."
Equations~(\ref{f_n}) and (\ref{g_n}) may be written as coupled
partial differential equations for the functions $f(p,l)=\sum_n
f_n(l)p^n$ and $g(p,l)=\sum_n g_n(l)p^n$:
\begin{eqnarray}
    \partial_l f &=& 2 f - \frac{1}{2}\tilde S\, g\, \partial^2_p f\,, \label{f}\\
    \partial_l g &=& \epsilon g - \frac{1}{2}\tilde S\, g\, \partial^2_p\, g \,,\label{g}
\end{eqnarray}
For $\epsilon>0$, Eq.~(\ref{g}) predicts the nontrivial stable
fixed--point polynomial $g^*(p)$, satisfying $\partial^2_p\, g^* =
2\epsilon/\tilde S$. In view of Eq.~(\ref{pm1}) the proper solution
is
\begin{equation}
     \label{gstar}
    g^*(p) = \frac{\epsilon}{\tilde S}\,(p -1)(p+\kappa) \,,
\end{equation}
where the parameter $\kappa$ is not  specified at this stage.
Substituting this into Eq.~(\ref{f}) one finds
\begin{equation}\label{asym_f}
    \partial_l f = 2f - {\epsilon\over 2}\,(p -1)(p+\kappa)\,\partial^2_p f \,.
\end{equation}
Since this is a linear  equation, one may look for its solution in
the form
\begin{equation}\label{asym_sol}
    f(p,l) = \sum_n  e^{\left (2 - E_n \right )l} \phi_n(p) \,,
\end{equation}
where the eigenfunctions $\phi_n(p)$ are solutions of the
stationary equation
\begin{equation}\label{eigen}
    E_n \phi_n(p) = {\epsilon\over 2}\, (p -1)(p+\kappa)\,\partial^2_p \phi_n(p) \, .
\end{equation}
Notice that if one chooses $\phi_n(j)$ to be a $n$-th degree
polynomial, the RHS of Eq.~(\ref{eigen}) is also a polynomial
of the same degree. It is clear then that one may always find a
solution $\phi_n(p)$ as a polynomial of the degree $n$. This means
that if one started from a polynomial of some degree $N$,
higher powers will not be generated by the RG. That is, the sum in
Eq.~(\ref{asym_sol}) is always confined to $1\leq n\leq N$.

To find the eigenenergies $E_n$ one needs to compare the coefficients
of the leading power of $p$ on both sides of Eq.~(\ref{eigen}).
This leads to $E_n = n(n-1)\epsilon/2$. From here and
Eq.~(\ref{asym_sol}) one finds that the  scaling dimensions of the
coefficients of the $f(p)$ polynomial {\em at the nontrivial 
fixed--point}, Eq.~(\ref{gstar}), are
\begin{equation}\label{fnscaling}
  [f_n]= 2-E_n=2-n(n-1)\,\epsilon/2\, .
\end{equation}
The conclusion is that it is sufficient to keep the polynomials
$g(p)$ to be of the second degree [cf. Eq.~(\ref{gstar})] while
$f(p)$ to be of the lowest possible degree consistent with the
symmetries of the model.

In the PC model of Sec.~\ref{secPC}, $g(p)$ must be even and thus
$\kappa =1$, leading to $g(p)=g_2(p^2-1)$. On the other hand, $f(p)$
must be odd and at least of degree $n=3$ [$n=1$ odd polynomial
can not satisfy Eq.~(\ref{pm1})]. Therefore it can be chosen  to  be
$f(p)=f_3(p^3-p)$. At the fixed point $g_2^*=\epsilon/\tilde S$ and
$[f_3]=2-3\epsilon$. This justifies treatment of Sec.~\ref{secPC}
with the identification $v=g_2$, and $u=f_3$.

In the reversible model $A\leftrightarrow 2A$ of
section~\ref{sec_rev} $g(p)=v(p^2-p)$ and thus $\kappa=0$. On the
other hand, $f(p)=u(p^2-p)$ which is the proper stationary
eigenfunction of Eq.~(\ref{eigen}): $\phi_2(p)$. As a result, no
other terms in $f(p)$ polynomial are generated upon
renormalization. This is consistent with the robustness of the
rectangular structure. Since $u=f_2$, its scaling dimension
according to Eq.~(\ref{fnscaling}) is $[u]=2-\epsilon$.



\begin{thebibliography}{xxx}
\bibitem{foot1}In fact the action, Eq.~(\ref{Reggeon}), after a proper rescaling
possesses the rapidity symmetry $p \rightarrow -q$, $q \rightarrow
-p$, and $t \rightarrow -t$, which shows the equivalence of $u$ and
$v$ vertices. This ``rapidity reversal'' symmetry of the DP action
implies the asymptotic equivalence of the survival probability and
the order parameter decay at criticality. Due to this symmetry, the
number of independent critical exponents for directed percolation
is reduced compared to general absorbing phase transitions (see
Refs.~[\onlinecite{Odor04,Lubeck04}] and references therein).
\bibitem{foot2}In principle, $k=4$ may exhibit a marginally
non-mean-field behavior in $d=1$, but we shall not try to
investigate it here.
\bibitem{footnew} This leaves the possibility that even for
$d>2$ the RG flows towards a strong--coupling fixed point, if the
bare nonlinearity is sufficiently strong. As a result one may
expect to observe a non-mean-field behavior for $d>2$.
\end{thebibliography}
\end{document}